\documentclass[12pt]{article}
\usepackage{latexsym}
\usepackage{amssymb}
\title{\vspace{.8in}\LARGE \bf Solving the Hierarchy Problem\\ with
  Noncompact Extra Dimensions\vspace{.3in}} 

\author{Andrew G. Cohen$^a$ and David B. Kaplan$^b$\ \thanks{\tt
    cohen@bu.edu, dbkaplan@phys.washington.edu}\\ \\  
  \small \it  $^a$Department of Physics, Boston University, Boston, MA
  02215, USA\\  
  \small \it $^b$ Institute for Nuclear Theory, 1550, University of
  Washington, Seattle, WA 98195-1550, USA \\ \\ 
}

\date{October 16, 1999}

\begin{document}

\newcommand\CL{{\cal L}}
\newcommand\CO{{\cal O}}
\newcommand\half{\frac{1}{2}}
\newcommand\beq{\begin{eqnarray}}
\newcommand\eeq{\end{eqnarray}}
\newcommand\eqn[1]{\label{eq:#1}}
\newcommand\intg{\int\,\sqrt{-g}\,}
\newcommand\refeq[1]{eq. (\ref{eq:#1})}

\begin{titlepage}
  \maketitle
  \begin{picture}(0,0)(0,0)
    \put(350,350){BU-HEP-BUHEP-99-26}
    \put(350,335){DOE/ER/40561-76-INT99}
  \end{picture}
  \begin{abstract}
    We show that gravitational effects of global cosmic 3-branes can
    be responsible for compactification from six to four space-time
    dimensions, naturally producing the observed hierarchy between
    electroweak and gravitational forces. The finite radius of the
    transverse dimensions follows from Einstein's equation, and is
    exponentially large compared with the scales associated with the
    3-brane. The space-time ends on a mild naked singularity at the
    boundary of the transverse dimensions; nevertheless unitary
    boundary conditions render the singularity harmless.
  \end{abstract}
  \thispagestyle{empty}
\end{titlepage}

\section{Introduction}

The possibility that we are living in more than four dimensions has
been considered sporadically by physicists since the time of
Kaluza \cite{Kaluza:1921} and Klein \cite{Klein:1926a,Klein:1926b}.
Although originally motivated by the possible unification of gravity
with gauge forces,  more 
recently there have been speculations that extra dimensions may also
play an important role in addressing several other outstanding
phenomenological problems --- in particular, the smallness of the
cosmological constant, the origin of the observed hierarchy
between the $W$ mass and the Planck scale, the nature of flavor, the
possible source of supersymmetry breaking, {\it etc.}  In this paper
we focus on the possibility that extra dimensions may provide a
resolution of the hierarchy problem.

There have been two distinct approaches using extra dimensions to deal
with the hierarchy problem.  One relies on compact extra dimensions,
curled up in such a way that 4-d phenomenology prevails.  The other
uses noncompact extra dimensions of infinite extent.  The first
category includes the ``large extra dimension'' approach, which
postulates a fundamental scale $\Lambda \sim 10-100$ TeV along with
Kaluza-Klein compactification at a large radius $R$ ($\sim$
millimeters for two extra dimensions).  The effective 4-d Planck scale
is then the fundamental scale $\Lambda$ times powers of $\Lambda R$
\cite{Antoniadis:1990ew,Arkani-Hamed:1998rs,Antoniadis:1998ig,Sundrum:1998ns,Arkani-Hamed:1998kx}.
If  $R$ is much larger than $\Lambda^{-1}$, the effective Planck scale
will be much larger than the fundamental scale. This does not solve
the hierarchy problem but rather reformulates it into a dynamical
question: why is the radius $R$ of the extra dimensions so much larger
than $1/\Lambda$? Mechanisms exist which can answer this question
\cite{Arkani-Hamed:1999pc}. Another possible hierarchy solution
arising from compact extra dimensions is the model of
ref. \cite{Randall:1999ee}, where an exponential warp factor gives
rise to the observed hierarchy, requiring an extra dimension only one
or two orders of magnitude larger than the fundamental length scale. A
dynamical mechanism to stabilize the radius is still needed, such as
first  suggested in \cite{Goldberger:1999uk,DeWolfe:1999cp}.

It has been shown that ordinary 4-d gravity can result from a theory
with infinite extra dimensions, provided that  there exists a
graviton bound to our 4-d submanifold
\cite{Visser:1985qm,Randall:1999vf,Gogberashvili:1999ad}. Such models
do not in general solve the hierarchy problem. In
refs. \cite{Arkani-Hamed:1999hk,Lykken:1999nb} it was suggested that, since the wave
function of the bound 4-d graviton falls off exponentially in the
direction transverse to the 3-brane that binds it, the observed
hierarchy of forces may be due to a displacement of our world from
that 3-brane.

In this paper we offer an alternative where the hierarchy arises from
finite, but noncompact extra dimensions.  The phenomenology closely
resembles the large extra dimension scenario
\cite{Arkani-Hamed:1998rs,Antoniadis:1998ig,Arkani-Hamed:1998kx}, but
provides a dynamical explanation for the large radius. However, unlike
these earlier scenarios, the extra dimensions we consider form a
non-compact space of exponentially large proper size, terminating at a
singularity.  The spacetime we analyze is the solution of Einstein's
equation in the presence of a global cosmic string
\cite{Cohen:1988sg}.

We begin by solving Einstein's equation for a global cosmic ``string''
in $d+2$ dimensions. By a ``string'' we will always mean a
$(d-1)$-brane with 2 transverse spatial dimensions.  We analyze scalar
and gravitational waves in this background metric, and demonstrate
that although the spacetime includes a
naked singularity, it is sufficiently mild that unitary boundary
conditions can be applied. We then display the hierarchy between
gravitational and gauge forces, showing that it is not the result of
any fine tuning, and does not lead to disagreement with gravitational
force experiments. We conclude by discussing possible generalizations
of our example.

\section{The metric about a global string}
\label{sec:metric}

A global cosmic string, or vortex, is a topologically stable scalar
field configuration with nontrivial homotopy for some internal
symmetry manifold.  The simplest example arises in a scalar field
theory with a spontaneously broken global $U(1)$ symmetry.  If the
expectation value of the scalar field in the ground state is
$\langle\Phi\rangle=f$, a vortex with unit winding number corresponds
to the field configuration $\Phi(r) = f(r) e^{i\theta}$, where
$\mathop{\lim}\limits_{r\to\infty}f(r) = f$. Unlike a vortex with the
$U(1)$ gauged (such as the Abelian Higgs model), a global vortex
has nonzero energy density outside its core, falling off as $1/r^2$.
It is therefore not surprising that, when gravity is included, a
curvature singularity appears exponentially far from the core of the
vortex, as was found in earlier work by the present authors for
vortices in four spacetime dimensions \cite{Cohen:1988sg}.

Generalization of this earlier work to branes in $d+2$ spacetime
dimensions is straightforward. We assume that outside the core of a
$(d-1)$-brane, the stress-energy tensor is that of a charged scalar
field with Lagrange density\footnote{We adopt the mostly plus sign
convention.}  $\CL =
-g^{\mu\nu}\partial_\mu\Phi^*\partial_\nu\Phi-V(\Phi)$ and expectation
value
\begin{eqnarray}
  \Phi= f^{d/2} e^{i\theta}\ , \eqn{vort} 
\end{eqnarray} 
where $f$ has dimensions
of mass.  The scalar field \refeq{vort} is assumed to minimize the
scalar potential $V(\Phi)$, and we tune the bulk cosmological constant
to zero.  We therefore look for a metric which has a $d$-dimensional
Poincar\'e invariant ``longitudinal'' space, and rotational invariance
in the transverse plane. The most general metric of this kind may be
put in the form
\begin{equation}
  \label{eq:umetric}
  ds^2 = {\bar A}(u)^2 \eta_{ab} dx^a dx^b + \gamma^2 {\bar B}(u)^2
  (du^2 + d\theta^2)\ , 
\end{equation}
where $x^a$ parameterizes the longitudinal $d$-dimensional space,
$\eta_{ab}$ is the flat Minkowski metric, and $\{u,\theta\}$ are the
transverse coordinates, with $\theta\in [0,2\pi)$.  The parameter
$\gamma$ has dimensions of length, while $u$, $\theta$ and the
functions $\bar A$ and $\bar B$ are dimensionless.  We will make the
somewhat perverse choice of placing the string core at large values of
$u$, while the singularity will appear at $u=0$.

The non-zero components of the stress-energy tensor which follow from
$\cal L$ and \refeq{vort} are:
\begin{eqnarray}
  \label{eq:stress}
  T^i_j = \delta^i_j \frac{f^d}{\gamma^2 {\bar B}^2} \ ,\qquad
  T^u_u = \frac{f^d}{\gamma^2 {\bar B}^2} \ ,\qquad
  T^\theta_\theta = -\frac{f^d}{\gamma^2 {\bar B}^2} \ .
\end{eqnarray}
The Einstein equations for this system are 
\begin{equation}
  G^\mu_\nu = -\frac{1}{M_{d+2}^d} T^\mu_\nu
\end{equation}
where $M_{d+2}$ is the analogue of the Planck mass in $d+2$ dimensions.

On computing the Einstein tensor from the metric in \refeq{umetric},
we arrive at the three differential equations
\begin{equation}
  \begin{array}{rcl}
    (d-1)({\bar A}'\,)^2 + {\bar A} {\bar A}''&=& 0 \\ \\ 
    {\bar A} ({\bar B}'\,)^2 + {\bar B}\left[d\, {\bar A}' {\bar
        B}'-d\, {\bar B} {\bar A}'' -{\bar A} {\bar B}''\right]&=& 0
    \\ \\    
    {{\bar A} ({\bar B}'\,)^2 - {\bar B}\left[ d\, {\bar A}' {\bar B}'
        + {\bar A} {\bar B}''\right]  } &=& \frac{2}{u_0} {\bar A}
    {\bar B}^2 
  \end{array}    
\end{equation}   
with solution
\begin{eqnarray}
  \label{eq:AB}
  {\bar A}(u) = \left(\frac{u}{u_0}\right)^{1/d}\ , \qquad
  {\bar B}(u) = \left(\frac{u_0}{u}\right)^{(d-1)/2d} e^{(u_0^2 -
    u^2)/2u_0}\ . 
\end{eqnarray}
where
\begin{eqnarray}
  u_0 \equiv \left(\frac{M_{d+2}}{f}\right)^d\ .
\end{eqnarray}
For a complete exterior solution these functions must be matched onto
a metric valid within the vortex core, determining the parameter
$\gamma$, which naturally is of size $\gamma\sim 1/f$.

Note that our solution has a genuine curvature singularity at
$u=0$. (For example, the Laplacian of the Ricci scalar diverges as
$\sim u^{-2/d}$ at $u=0$.)  Furthermore, this is a naked singularity,
located at a finite proper distance from the string core near $u
\sim u_0$.

Because of the existence of a naked singularity it is not immediately
clear whether or not sensible physical results follow from our solution
\cite{Gregory:1988xc,Gibbons:1988pe,Gregory:1996dd}. 
There are two different problems that might arise.  The first is that
the existence of the singularity is not reliable, precisely because
gravitational forces become strong there. A higher derivative
correction to the Einstein-Hilbert action for example, would probably
eliminate---or at least change---the nature of the singularity. We will
argue later that, while this may indeed happen in extensions of
Einstein's theory, it would likely not change the conclusions we
arrive at by assuming the singularity to be physical.

A second problem is that the conservation of energy and momentum might
be violated at the singularity.  This can be phrased as a question
concerning the boundary conditions that must be applied to fields at
the singularity \cite{Gell-Mann:1985if}.  Just as the irregular
Legendre functions which appear when solving the non-relativistic
Schr\"odinger equation for the hydrogen atom must be discarded, as
they correspond to probability loss at the origin, so must any
solution here which loses energy or momentum through the
singularity. Thus we must search for conditions on fields propagating
in the background metric that insure conservation of quantum numbers
at the singular boundary of the spacetime. The ability to find such
conditions is not guaranteed if the singularity is too strong,
however.  To address this issue we first consider propagation of a
massless scalar in our background metric; we then consider
gravitational perturbations.

\section{Scalar fields and boundary conditions}
\label{sec:scalars}

We consider a  massless scalar field $\phi$ satisfying the wave
equation
\begin{equation}
  \label{eq:scalarwave}
  \Box\phi= \frac{1}{\sqrt{-g}} \partial_\mu (\sqrt{-g}g^{\mu\nu}
  \partial_\nu \phi) = 0\ .
\end{equation}
It is convenient to introduce a second parameterization of the
transverse radial dimension:
\begin{equation}
  \label{eq:zmetric}
  ds^2 = A(z)^2 \eta_{ab} dx^a dx^b +  A(z)^2 dz^2 + \gamma^2 B(z)^2
  d\theta^2\ ,
\end{equation}
with
\begin{equation}
  \label{eq:utoz}
  z = \gamma \int_0^u \frac{{\bar B}(u')}{{\bar A}(u')} \, du'\
  ,\qquad  A(z) \equiv {\bar A}(u(z))\ ,\qquad  B(z) \equiv {\bar
    B}(u(z))\ .  
\end{equation}
The functions ${\bar A}(u)$ and ${\bar B}(u)$ are the solutions found
previously in \refeq{AB}. For small $u$, $z$ behaves like $z\sim
u^{(d-1)/2d}$, so that the singularity at $u=0$ is also at $z=0$. Near
the singularity
\begin{eqnarray}
  \eqn{singularAB}
  A(z) \to z^{2/(d-1)}\ ,\qquad
  B(z) \to \frac{1}{z} \ .
\end{eqnarray}

Using the isometries of the background metric, we search for solutions
of the form  
\begin{eqnarray}
  \label{eq:varphi}
  \phi = e^{ip\cdot x} e^{i n \theta}\frac{\varphi(z)}{\psi_1(z)}\
  ,\qquad \eta_{ab}p^ap^b = -m^2\ , 
\end{eqnarray}
where $p\cdot x \equiv \eta_{ab} p^a x^b$ and
\begin{eqnarray}
  \psi_1(z)=(g^{zz}\sqrt{-g})^{1/2} = (\gamma A^{d-1} B)^{1/2}\ .
  \eqn{psi1}
\end{eqnarray}
The parameter $m$ will be the conventional $d$-dimensional mass of
this mode.  The function $\varphi(z)$ then satisfies the equation
\begin{eqnarray}
  \left[-\frac{{\rm d}^2\phantom{z}}{{\rm d}z^2}+ V(z) +n^2\frac{
      A^2}{ B^2}\right]
  \varphi(z) = m^2 \varphi(z)\ ,\qquad
  V(z) = \frac{\psi_1''}{\psi_1}
\end{eqnarray}
This equation can be recast in the more convenient form
\begin{eqnarray}
  \label{eq:wave}
  \left[ {\bar Q} Q +n^2\frac{
      A^2}{ B^2}\right]
  \varphi(z) = m^2 \varphi(z)
\end{eqnarray}
where 
\begin{eqnarray}
  Q = -\frac{{\rm d}\phantom{z}}{{\rm d}z} + \frac{d
    \log\psi_1(z)}{dz}\ ,\qquad {\bar Q}= \frac{{\rm
      d}\phantom{z}}{{\rm d}z} + \frac{d \log\psi_1(z)}{dz}\ . 
\end{eqnarray}
In this form, the scalar wave equation clearly has two zeromode
solutions with $n=m=0$. One such solution is easily identified
\begin{eqnarray}
  \varphi(z) = \psi_1(z)\ .
\end{eqnarray}
The second zeromode follows as
\begin{eqnarray}
  \varphi(z) = \psi_2(z) \equiv\psi_1(z) \int^z
  \frac{dz'}{\psi_1(z')^2} = \psi_1(z) \log(u(z))\ . 
\end{eqnarray}
Near the singularity at $z=0$ these two solutions behave as
\begin{eqnarray}
  \label{eq:singles}
  \psi_1(z)\to \sqrt{z}, \qquad \psi_2(z) \to \sqrt{z}\log z\ .
\end{eqnarray}
Eq. (\ref{eq:wave}) shows that the $m^2$ and $n^2$ terms become
irrelevant near the singularity, and thus {\em all} solutions behave
similarly to the two zeromode solutions in the limit $z\to 0$. We will
refer to the solutions behaving as $\sqrt{z}$ near $z=0$ as the ``regular''
solutions, while those behaving as $\sqrt{z}\ln z$ for small $z$ will
be called the ``irregular'' solutions.

The fact that our spacetime is geodesically incomplete should not
matter provided that no conserved quantities are allowed to leak out
through the boundary. The $d$-dimensional Poincar\'e isometries of the
metric \refeq{umetric}, as well as the axial rotation isometry, lead
to conservation laws. The Killing vectors corresponding to these
isometries are
\begin{eqnarray}
  \xi^{(a)}_\mu &=& A^2 \delta^a_\mu \nonumber \\
  \xi^{(a b)}_\mu &=& A^2 \left( \delta^a_c\delta^b_\mu -
    \delta^a_\mu\delta^b_c\right) x^c\\  
  \xi^{(\theta)}_\mu &=& B^2\, \delta_\mu^\theta\nonumber
\end{eqnarray}
with $a,b,c=1\ldots d$. For each of these Killing vectors we may
contract with the stress-energy tensor to form a current
\begin{eqnarray}
  J^\mu = T^{\mu\nu} \xi_\nu
\end{eqnarray}
satisfying a covariant conservation law
\begin{eqnarray}
  \frac{1}{\sqrt{-g}} \partial_\mu (\sqrt{-g} J^\mu) = 0 \ .
\end{eqnarray}
Therefore we demand that the flux through the singular boundary of our
spacetime must vanish for each of these currents. For example, for the
translation isometries this requires
\begin{eqnarray}
  \label{eq:bc}
  \lim_{z\to 0} \sqrt{-g} J_{(a)}^z =\lim_{z\to 0}  \sqrt{-g} g^{zz}
  \partial_a \phi \partial_z\phi = 0 \ .
\end{eqnarray}
Using \refeq{singles} and the form of $\phi$ \refeq{varphi} we see
that  the flux through the singularity for the regular solutions
vanishes, while that for the irregular solutions diverges. Therefore
the boundary conditions boundary conditions we have chosen eliminate the irregular
solutions. The other currents lead to the same condition.

Once the irregular solutions have been discarded, it follows that that
$\bar Q = Q^\dagger$; that is, $(\varphi_1,\bar Q \varphi_2) = (Q
\varphi_1, \varphi_2)$ where $\varphi_{1,2}$ are arbitrary regular
solutions, since the difference between the two expressions is a
vanishing surface term.  Therefore $\bar Q Q$ is a positive
semidefinite operator, implying that the $m^2$ eigenvalues are all
positive, so that no tachyonic modes are seen in the $d$-dimensional
world.

In fact the  boundary condition \refeq{bc} (supplemented by
appropriate boundary conditions near the string core) 
is precisely the condition necessary to insure that the eigenvalue
problem specified by eqs. (\ref{eq:scalarwave},\ref{eq:varphi}) is of
the classic Sturm-Liouville type. Thus not only is the spectrum
non-tachyonic, but the eigenfunctions are also  complete, discrete,
and my be chosen to be orthogonal.

\section{Gravitational perturbations and Newton's law of gravitation}
\label{sec:grav}

We turn to the metric fluctuations about the solution \refeq{AB}.
The metric in $(d+2)$-dimensions has $(d+2)(d-1)/2$ physical
polarizations; for a $d$-dimensional observer these decompose into a
$d$-dimensional graviton, 2 $d$-dimensional vectors and 3
$d$-dimensional scalars. We first focus on the $d$-dimensional
graviton fluctuations.  We consider a perturbation of the metric
of the form
\begin{equation}
  \label{eq:zflucmetric}
  ds^2 = A(z)^2 (\eta_{ab}+ h_{ab}) dx^a dx^b +  A(z)^2 dz^2 +
  \gamma^2 B(z)^2 d\theta^2\ ,
\end{equation}
and impose the gauge conditions $\partial_a h^{ab} = h^a_a = 0$,
following the analysis of ref. \cite{Randall:1999vf}. Using the
isometries of the background metric we will search for solutions of
the form
\begin{equation}
  \label{eq:metricfluc}
  h_{ab} = \varepsilon_{ab} e^{i p\cdot x} e^{i n \theta}
  \frac{\varphi(z)}{\psi_1(z)}\ , 
\end{equation}
where $\epsilon_{ab}$ is a constant polarization tensor, and $\psi_1$
is the same function\footnote{Note that, strictly speaking, it is
the combination $(\sqrt{-g}g^{ii})^{1/2}$ that appears in the
denominator of \refeq{metricfluc}. For our coordinate choice, in which 
$g^{ii} = g^{zz}=1/A^2$, this equals  $\psi_1$.} defined in \refeq{psi1}.

We can already identify one solution of the form \refeq{metricfluc}:
the original metric \refeq{zmetric} has an invariance under restricted
general coordinate transformations along the brane. This is enough to
insure the presence of a massless graviton along the brane. The
fluctuation in the metric corresponding to this mode may be found by
replacing $\eta_{ab} \to {\bar g}_{ab}$; that is, the wave function of
this mode is just an $h_{ab}$ independent of the transverse
coordinates:
\begin{equation}
  \label{eq:zeromode}
  \varphi(z) = \psi_1(z)\ ,
\end{equation}
with $n=m=0$. 

The coupling of this mode to stress-energy along the brane may be
computed by examining the action for this fluctuation. Using
eqs. (\ref{eq:metricfluc}, \ref{eq:zeromode}) we get an action
\begin{equation}
  \label{eq:effaction}
  S = -M_{d+2}^d  \int  \sqrt{-g} \, dz\, d\theta \, d^d\!x
  \sqrt{-(\eta+h)} \frac{R_d}{A^2}
\end{equation}
where  $R_d$ is the $d$-dimensional curvature
computed from $(\eta+h)_{ab}$. This allows identification of the
$d$-dimensional  Planck scale as \footnote{In principle we only trust
  the integrand in the region outside the string core. However the
  integral is dominated by this region---inclusion of the string core
  changes the value of the integral by an exponentially small amount.}:
\begin{equation}
  \label{eq:newton}
  M_d^{d-2} = M_{d+2}^d \, \int \psi_1(z)^2 \,dz\,d\theta = e^{u_0}
  u_0^{(d+1)/2d} \Gamma\left(\frac{d-1}{2d}\right) \pi \gamma^2 M^d_{d+2}\ .
\end{equation}
This form demonstrates the connection between the gravitational
coupling in $d$-dimensions, and the normalization of the zeromode
$\psi_1$.  The factor $e^{u_0}$ dominates this expression, providing
an exponential hierarchy. For the especially interesting case of
$d=4$, with $\gamma=1/f$, a ratio of $M_6/f \simeq 2.7$ yields a
hierarchy for $M_4/M_6$ of over $10^{17}$!

The Einstein equation for general fluctuations of the form
\refeq{metricfluc} reduces to the scalar wave equation \refeq{wave}
which we considered in the previous section:
\begin{equation}
  \label{eq:einsteinfluc}
  \left[ {\bar Q} Q +n^2\frac{
      A^2}{ B^2}\right]
  \varphi(z) = m^2 \varphi(z)
\end{equation}
where $Q$ is the same operator as in the scalar case.  Our analysis is
therefore identical to that of the previous section: we discard all
solutions for which $\varphi(z)$ behaves like $\sqrt{z} \log z$ near
the singularity. The remaining solutions include the zeromode,
behaving as a massless $d$-dimensional graviton on the
brane\footnote{Unlike in the scalar case, we are assured that there
will be a solution with $m^2=0$, since whatever occurs in the core of
the vortex will maintain general coordinate invariance.}, and the
Kaluza-Klein like modes with non-zero $m^2$. All these modes have
$\varphi(z)$ vanishing like $\sqrt{z}$ as we approach the
singularity. Since $\bar Q Q$ is positive semidefinite, we are assured
that the Kaluza-Klein modes all have real masses, and that there are
no gravitational instabilities. Furthermore, the spectrum is discrete,
with spacing set by
\begin{eqnarray}
  \label{eq:deltam}
  \Delta m^2 \sim 1/z_0^2\sim  
  \frac{(M_{d+2})^{d}}{( M_{d})^{d-2}}\ , 
\end{eqnarray}
where $z_0=z(u_0)$ and we made use of \refeq{newton}. This is the same
relation one expects in the large extra dimension scenario.  Since
these Kaluza-Klein modes are coupled with gravitational strength, they
are still acceptable, even if they are extremely light
\cite{Arkani-Hamed:1998rs}.

As for the other modes of the graviton field, we expect the three
scalar modes and one of the vector modes that arise from the
decomposition to have masses on the same scale as $\Delta m$.
However, one of the vector modes is corresponds to the rotational
isometry of the metric.  Normally, one would then expect this field to
be the vector potential associated with an unbroken $U(1)$ gauge
symmetry in the $d$-dimensional world.  However, the scalar field
$\Phi$ breaks the rotational symmetry.  Therefore this gravi-photon
eats the Goldstone boson associated with phase
rotations of $\Phi$, and gets a mass$^2$ comparable to the $\Delta m^2$ 
of \refeq{deltam}.

\section{Conclusions and speculations}
\label{sec:conc}

We have shown that the metric outside a global cosmic 3-brane provides
a dynamical determination of the hierarchy between the high
scale associated with Newton's constant, $10^{19}$ GeV, and the low
scale associated with particle physics, a few TeV.
With no fine tuning of fundamental parameters, this metric
produces a finite, large transverse area. Small fluctuations about
this background metric behave as a normal, 4-d graviton along the
brane. The coupling of this mode to excitations on the brane, the
effective Newton constant, is the 6-d Newton constant, taken to be
characterized by the TeV scale\footnote{This scale may need to be more
nearly 100 TeV to satisfy all phenomenological constraints
\cite{Hall:1999mk}.}, divided by the area of the transverse
dimensions. Since this area is related to the intrinsic scale by a
factor $\sim \exp((M_6/f)^4)$, a ratio of $M_6/f\sim 2.7$ easily
provides the necessary hierarchy.

The fact that our metric is geodesically incomplete does not make our
solution inconsistent, since we are able to find boundary conditions
on our fields that guarantee no conserved quantities are lost
through the singularity.  Nevertheless, we expect that physics beyond
Einstein's theory will alter the nature of the singularity, and
possibly eliminate it altogether. Even in the latter case, gravity would
become strong, even if not singular, only at an exponentially large
distance from the vortex. The key feature of our solution --- the dynamical
generation of an exponentially large length scale --- would remain
intact.  Furthermore, it is quite plausible that the low lying
solutions to the wave equations we have considered will not be
appreciably altered, precisely because the boundary conditions we have
chosen make the solutions insensitive to what is happening at the
singularity. They would be analogous to the low lying quantum
mechanical states in a deep square well potential, which are
insensitive to whether the potential is actually infinite outside the
box, or whether it is finite but large.

In our discussion the nature of the 3-brane played little role. The
only requirement was the existence of stress-energy of the form
eq. (\ref{eq:stress}). Such a stress-energy tensor applies when scalar
field equations have stable brane solutions from spontaneously broken
global symmetries (such as a cosmic ``string''), or from long range
fieldss~\cite{Dvali:1999tq} attached to D-brane\footnote{The example
  of Arkani-Hamed, 
Dimopolous and March-Russell \cite{Arkani-Hamed:1999yp} for obtaining
logarithmic running through a massless scalar field coupled to a
3-brane provides another example of a D-brane as a source for a massless
scalar field}. While we specifically discussed a brane with
codimension equal to two, it seems possible that one could find
similar solutions with exponentially large extra dimensions starting
with branes of other codimensions --- preferably a D-brane that could
support a realistic 4-d world.  
It is interesting to speculate whether
the winding number of a bulk topological field about this brane could
determine the number and chirality of families living on the brane.

\bigskip
\noindent\medskip\centerline{\bf Acknowledgments}

We thank Nima Arkani-Hamed, Ann Nelson, Lisa Randall, Martin Schmaltz, 
and Raman Sundrum for useful conversations.
We would like to express our gratitude to the Aspen Center for
Physics, where this work was begun.
A.G.C. is supported in part by DOE grant
\#DE-FG02-91ER-40676; D.B.K. is supported in part by DOE grant
\#DOE-ER-40561.


\end{document}